\documentclass[prl,twocolumn,showpacs,preprintnumbers,amsmath,amssymb]{revtex4}
\usepackage{graphicx}
\usepackage{dcolumn}
\usepackage{bm}
\usepackage{epsfig}

\begin{document}

\input{epsf}

\title{Coherent dynamics of domain formation in the Bose Ferromagnet}
\author{Qiang Gu$^{1,2}$, Haibo Qiu$^{1,3}$}
\affiliation{$^1$Department of Physics, University of Science and
Technology Beijing, Beijing 100083, China\\$^2$Institut f\"ur
Laser-Physik, Universit\"at Hamburg, Luruper Chaussee 149, 22761
Hamburg, Germany\\$^3$Institute of Theoretical Physics, Lanzhou
University, Lanzhou 730000, China}

\date{\today}
\textbf{}

\begin{abstract}
We present a theory to describe domain formation observed very
recently in a quenched $^{87}$Rb gas, a typical ferromagnetic spinor
Bose system. An overlap factor is introduced to characterize the
symmetry breaking of $M_F=\pm 1$ components for the $F=1$
ferromagnetic condensate. We demonstrate that the domain formation
is a co-effect of the quantum coherence and the thermal relaxation.
A thermally enhanced quantum-oscillation is observed during the
dynamical process of the domain formation. And the spatial
separation of domains leads to significant decay of the $M_F=0$
component fraction in an initial $M_F=0$ condensate.
\end{abstract}


\maketitle

Very recently, the Berkeley group observed spontaneous symmetry
breaking in $^{87}$Rb spinor condensates~\cite{Berkeley}.
Ferromagnetic domains and domain walls were clearly shown using an
in-situ phase-contrast imaging. This appears the first image of the
domain structure in a Bose ferromagnet. Although ferromagnetism has
been intensively studied in the context of condensed matter physics
and is regarded as one of the best understood phenomena in
nature~\cite{Mohn}, the description of ferromagnetism is not yet
complete. The conventional ferromagnets being considered are usually
comprised of either classical particles (insulating ferromagnets) or
fermions (itinerant ferromagnets) while Bose systems are seldom
touched~\cite{Gu0}. The realization of cold spinor $^{87}$Rb
gases~\cite{Steng}, a typical ferromagnetic Bose system, has
provided an opportunity to study Bose ferromagnets and thus opens up
a way to a comprehensive understanding of ferromagnetism in all
kinds of condensed matters.

The ferromagnetic spinor Bose gas has attracted numerous theoretical
interests~\cite{Ho1,Ohmi,Gu1,Kisz,Isoshima,Zhang,Schma1,Chang1,Isoshima2,Zhang2,Mur}.
On one side, researchers expect that it will show some general
properties as conventional ferromagnets do. Ho~\cite{Ho1}, Ohmi and
Machida~\cite{Ohmi} pointed out that this system has a spontaneous
symmetry-broken ground state and a normal spin-wave excitations
spectrum at small wave vector ${\bf k}$, $\omega_s=c_s k^2$. On the
other side, researchers aim at exploring distinct features of the
system. Studies on thermodynamics and phase transitions have
revealed that the ferromagnetic spinor Bose gas displays a quite
surprising phase diagram. Its Curie point can be larger by
magnitudes than the energy scale of the ferromagnetic interaction
between bosons, and never below the Bose-Einstein condensation
point~\cite{Gu1,Kisz}. It means that once the Bose gas condenses, it
is already spontaneously magnetized.

A conventional ferromagnet usually has some domain structure below
the Curie point, as illustrated in Fig. 1a. But whether it is true
for a Bose ferromagnet is still somewhat controversial. One even
questions whether there exists a Curie point in ferromagntic Bose
gases~\cite{Isoshima,Zhang}, as the cold atomic gas under
experimental conditions is usually not in the thermodynamic limit
while the phase diagram mentioned above is derived from the
thermal-equilibrium grand canonical ensemble~\cite{Gu1,Kisz}.
Nevertheless, a number of theoretical works have discussed the
possibility of the domain formation~\cite{Gu0,Isoshima2,Zhang2,Mur}.
Within the mean-field theory, Zhang {\it et al.} found out that the
ferromagnetic condensate has a dynamical instability leading to
spontaneous domain formation in an initially magnetized Bose
gas~\cite{Zhang2}. Moreover, Mur-Petit {\it et al.} showed that a
multi-spin-domain structure manifests in $^{87}$Rb condensates at
finite temperatures\cite{Mur}. The Berkeley experiment confirms that
the Bose ferromagnet can indeed form domain structures at least
under certain conditions~\cite{Berkeley}. Then new questions come,
is the process of the domain formation similar to that inside a
conventional ferromagnet, and how does the domain formation affect
spin dynamics? We attempt to answer these questions in the present
letter.

\begin{figure}
\centering
\includegraphics[width=0.45\textwidth]{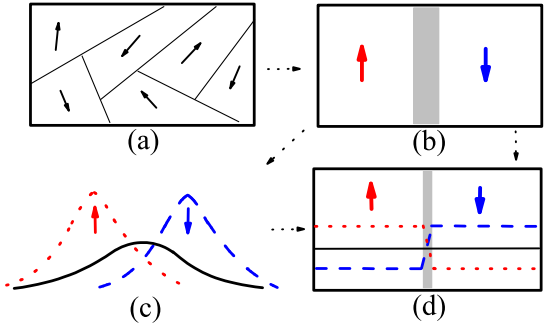}
\caption{ \label{fig:epsart1}Schematic domain structure inside a
Ferromagnet. (a) The bulk material shows no magnetism because
domains are randomly oriented. (b) The multi-domain configuration is
often simplified to a two-domain structure with opposite sign of
magnetization for theoretical convenience. The gray region denotes
the domain wall. (c) Two-domain structure for a ferromagnetic spinor
Bose-Einstein condensate. Dotted, dashed and solid lines represent
the normalized particle distributions of $M_F=1$, $M_F=-1$ and
$M_F=0$ bosons respectively. (d) In a homogeneous system, the
normalized particle distribution is taken as a constant in each
domain; for $M_F=0$ bosons it remains a constant through the whole
system. }
\end{figure}

We start with the Hamiltonian for an $F=1$ system in the following
form~\cite{Ho1}
\begin{eqnarray}\label{eq:1}
H &=& \int d^3 r \left[ \frac{\hbar^2}{2m} \nabla \psi_a^{\dagger}
     \cdot \nabla \psi_a
  - (\mu - U) \psi_a^{\dagger}\psi_a  \right.\nonumber \\
  &+& \left. \frac{g_0}{2}\psi_a^{\dagger}\psi_{a'}^{\dagger}\psi_{a'}\psi_a
  + \frac{g_2}{2}\psi_a^{\dagger}\psi_{a'}^{\dagger} {\bf F}_{ab}\cdot
  {\bf F}_{a'b'}\psi_{b'}\psi_b\right],
\end{eqnarray}
where $\psi_a({\bf r})$ is the field annihilation operator for an
atom in state $M_F=a$ at point ${\bf r}$, $\mu$ is the chemical
potential and $U$ is the trapping potential. $g_0$ and $g_2$ are the
spin-independent and spin-dependent mean-field interaction,
respectively. For macroscopically occupied Bose systems it is common
to replace the field annihilation operator for the $a$th spin
component by its expectation value, i.e. $\varphi_a({\bf r},t)
\equiv \langle \psi_a ({\bf r},t)\rangle $, which for spinor
condensates is conveniently expressed as
\begin{equation}\label{eq:2}
  \varphi_a({\bf r},t) = \sqrt{N_a(t) \eta_a ({\bf r},t)} e^{i\phi_a ({\bf r},t)}.
\end{equation}
Here $N_a(t)$ is the number of condensed particles, $\eta_a({\bf
r},t)$ denotes the normalized particle distribution with $\int
d^3r\eta_a=1$ and $\phi_a ({\bf r},t)$ the phase. The multiplier of
$N_a$ and $\eta_a$ refers to the condensed particle density of the
$a$-component, and the total particle density is $\rho({\bf
r},t)=N_{a}(t)\eta_{a}({\bf r},t)$. We suppose that the particle
distribution can be different for distinct components in our
approach. Neglecting excitations, the Hamiltonian is simplified to
$H = H_0 + H_u + H_s$, with
\begin{eqnarray}\label{eq:3}
H_0 &=& \int d^3 r \left[ \frac{\hbar^2}{2m} N_a(\nabla \sqrt{\eta_a})^2
      +\frac{\hbar^2}{2m} N_a\eta_a(\nabla \phi_a)^2  \right] ,\nonumber \\
H_u &=& \int d^3 r \left[(U-\mu)N_a\eta_a + \frac{g_0}{2} N_a\eta_a N_{a'}\eta_{a'}\right] ,\nonumber \\
H_s &=& \int d^3 r \frac{g_2}{2} \left[
     N_{+}^2\eta_{+}^2 + N_{-}^2\eta_{-}^2 - 2N_{+}N_{-}\eta_{+}\eta_{-} \right.\nonumber \\
    &&+  2N_{0}\eta_{0}( N_{+}\eta_{+} + N_{-}\eta_{-}) \nonumber \\
    &&+ \left. 4 N_{0}\sqrt{N_{+}N_{-}} \eta_{0}\sqrt{\eta_{+}\eta_{-}}
      {\rm cos}\theta \right],
\end{eqnarray}
where $\theta=\phi_{+}+\phi_{-}-2\phi_0$ is the relative phase.

In case that the ground state of the condensate is symmetry-broken,
certain magnetic domain structure is formed spontaneously. Within
each domain the atomic magnetic moments are aligned in a
preferential direction. A direct consequence of domain formation is
that the $M_F=1$ and $M_F=-1$ components are spatially separated, as
portrayed in Fig. 1c. The integral $\int d^3 r \eta_{+}\eta_{-}$
measures the extent of overlap between the two, which is called the
overlap factor hereinafter. In general, several ``overlap factors"
should be introduced to Eqs.(3), e.g., $\alpha_{0\pm} = V\int d^3 r
\eta_{\pm}^2$, $\alpha_{1} = V\int d^3 r \eta_{+}\eta_{-}$,
$\alpha_{2\pm} = V\int d^3 r\eta_{0}\eta_{\pm}$ and $\alpha_{3} =
V\int d^3 r\eta_{0}\sqrt{\eta_{+}\eta_{-}}$, where $V$ is the volume
of the system. Treating the relative phase $\theta$ as a spatially
independent constant as previous theory
did~\cite{Isoshima2,Zhang2,Mur,Schma1,Chang1}, the term of $H_s$ in
Eqs. (3) is rewritten as
\begin{eqnarray}\label{eq:4}
H_s &=& \frac{g_2}{2V}\left[ \alpha_{0+} N_{+}^2
      + \alpha_{0-}N_{-}^2 - 2\alpha_{1}N_{+}N_{-} \right.\nonumber \\
    &&+  2N_{0}(\alpha_{2+}N_{+} + \alpha_{2-}N_{-}) \nonumber \\
    &&+ \left. 4 \alpha_{3} N_{0}\sqrt{N_{+}N_{-}} {\rm cos}\theta \right] .
\end{eqnarray}
According to their definition, these overlap factors are not totally
independent from each other and the number of independent ones can
be further reduced. For simplicity, we consider a homogeneous spinor
Bose gas with a two-domain structure, as shown in Fig. 1d, and
neglect the domain wall. In this case, we derive, after some
integration and algebraic manipulation, that there is only {\sl one}
independent overlap factor and the above equation is reduced to
\begin{eqnarray}\label{eq:5}
H_s &=& \frac{g_2}{2V}\left[ (2-\alpha)(N_{+}^2 + N_{-}^2) - 2\alpha N_{+}N_{-} \right.\nonumber \\
     &+& \left. 2 N_{0}(N_{+}+ N_{-}) + 4\sqrt{\alpha}N_{0}\sqrt{N_{+}N_{-}}{\rm cos}\theta \right] ,
\end{eqnarray}
where the reduced overlap factor is just given by $\alpha=V\int d^3
r \eta_{+}\eta_{-}$. Our model allows the $\alpha$-factor to vary
from one to zero, corresponding to the case that the two components
are from thoroughly mixed to completely separated. For a homogeneous
system, the gradient term $H_0$ can be doped. The spin-irrelevant
term $H_u$ remains a constant since the total density distribution
is hardly responsive to the evolution of domain
structure~\cite{Berkeley,Isoshima2}, $\rho({\bf r},t)\approx
\rho({\bf r},0)$. Thus the dynamics of domain formation is only
determined by the Hamiltonian expressed in Eq. (\ref{eq:5}).

The Berkeley experiment considered a pure spinor condensate
initially prepared in the unmagnetized state. It is important to
emphasize that the total spin is conserved in an atomic quantum gas
under experimental conditions\cite{Chang2,Kronj,Schma2}. Therefore
the particle numbers of $M_F=1$ and $-1$ component are always equal,
$N_{+}= N_{-}$. Such a system is described by the Hamiltonian
\begin{eqnarray}\label{eq:6}
{\cal H}_s = -(1-\alpha)(1- n_{0})^2 - 2 n_{0}(1-n_{0}) \left[1 +
\sqrt{\alpha}{\rm cos}\theta \right] .
\end{eqnarray}
Here ${\cal H}_s=H_s/(N\vert{\frac{g_2}{2}\vert})$ with total
particle number $N=N_{+}+ N_{-}+N_0$; $n_a=N_a/N$ is the fraction of
$M_F=a$ component. $n_0$ and $\theta$ form a pair of conjugate
variables and their equations of motion are given by
\begin{subequations}\label{eq:7}\begin{eqnarray}\label{eq:7a}
\frac{\partial}{\partial t}n_{0}&=& -2\sqrt{\alpha} n_0 (1-n_0) {\rm sin}\theta ,\\
\frac{\partial}{\partial t} \theta &=& 2(1-\alpha)(1-n_0) \nonumber\\
      &&- 2(1-2n_0) \left[1 + \sqrt{\alpha}{\rm cos}\theta \right] .
\end{eqnarray}\end{subequations}
The population dynamics depicts the process of the domain formation;
and the magnetization of magnetic domains is defined as
$m=N_{+}\eta_{+}-N_{-}\eta_{-}$.

\begin{figure}
\centering
\includegraphics[width=0.4\textwidth]{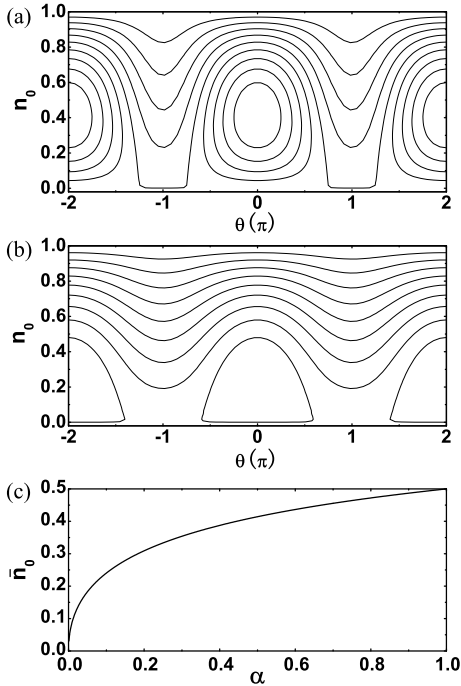}
\caption{ \label{fig:epsart2} Contour plot of the energy surface in
the $\theta$-$n_0$ plane with the overlap factor $\alpha=0.5$ (a)
and $\alpha=0.1$ (b). (c) shows the fraction of $M_F=0$ atoms at the
minima, $\bar{n}_0$, as a function of $\alpha$. }
\end{figure}

\begin{figure}
\centering
\includegraphics[width=0.4\textwidth]{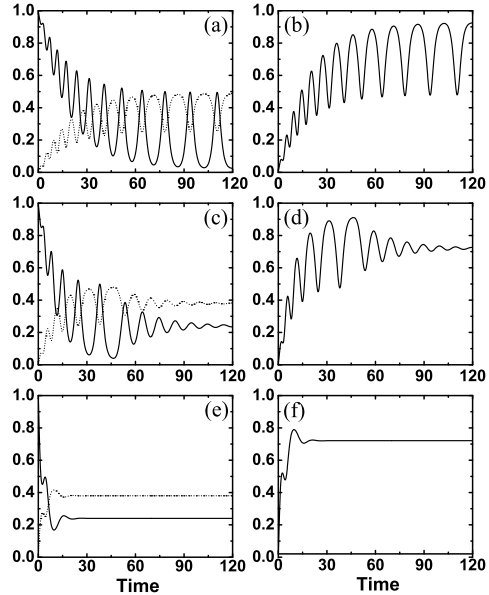}
\caption{ \label{fig:epsart3} Dynamics of domain formation for Bose
systems with the overlap factor $\alpha=0.1$ and $T_R=20$, $10$, and
$2$ from top to bottom. The left column shows the particle fractions
of $M_F=0$ ($n_0$, solid lines) and $M_F=\pm1$ ($n_{\pm}$, dotted
lines) components. The right column shows the magnetization of
magnetic domains, $m$. The initial state configuration is
($n_{+}=0.005,n_0=0.99,n_{-}=0.005$). }
\end{figure}

A number of theoretical works have investigated the spin dynamics of
the $^{87}$Rb spinor condensate~\cite{Zhang2,Schma1,Chang1}. We
notice that previous theories usually treated the three components
being mixed as they share the same spatial wave function, which is
known as the single-mode approximation. Therefore the domain
structure is smeared out and thus the ferromagnetic feature was not
sufficiently dealt with. Those theories correspond the special case
of $\alpha=1$ in this letter.

The dynamical behaviors of Eqs. (\ref{eq:7}) can be visualized by
the phase-space portrait with constant energy lines. Figure 2a and
2b plot the contour lines of energy with the overlap factor
$\alpha=0.5$ and $0.1$, respectively. In case of $\alpha=1$, there
are energy minima along $n_0=1/2$ at
$\theta=2n\pi$~\cite{Zhang2,Schma1,Chang1} and all the contour lines
are closed loops around those minima. As domains build up, the value
of $\alpha$ drops down and two apparent changes take place, seen in
Fig. 2a and 2b. (i) There exist two distinct regimes in the phase
space diagram. The newly appeared regime lies in the upper region of
the figures, which consists of a set of open curves. Each line
corresponds to a rotation type of solution, in which the relative
phase $\theta$ is "running" with the time. The closed orbits, lying
in the lower region, represent libration type of solutions, in which
$\theta$ oscillates around the minimum. (ii) The positions of those
minima move towards to smaller values of $n_0$. The minima points
are connected to the ground state of the system. Figure 2c shows the
$M_F=0$ particle fraction at the minima, $\bar{n}_0$. The less
$\alpha$ is, the smaller $\bar{n}_0$ is. It means that the domain
formation tends to reduce the number of $M_F=0$ atoms. This point
has also been affirmed by Mur-Petit {\it et al.} who obtained a
state with equipartition in populations, ($n_{+}\approx
1/3,n_0\approx 1/3,n_{-}\approx 1/3$), from a starting state
(0.005,0.99,0.005)~\cite{Mur}. We derive that $n_0=n_{+}=n_{-}=1/3$
when $\alpha$ drops to $0.25$. In the limit case of $\alpha=0$,
$n_0=0$ and $n_{+}=n_{-}=0.5$.

The state of the quenched $M_F=0$ condensate is viewed as a point
lying in the upper region of the phase space diagram (Fig. 2a or
2c), when Eqs. (\ref{eq:7}) yield a self-trapping
solution\cite{Smerzi}. This motion reflects the quantum mechanical
nature of the Bose-Einstein condensate. The oscillating amplitude of
$n_0$ is very small and the resulting magnetization $m$ is so low in
magnitude that it is hard to be probed experimentally during this
stage\cite{Berkeley}. The self-trapping effect prevents the growth
of magnetic domains. This case is similar to the classical Larmor
precession of a spin around magnetic field: it is rotating all the
time, but the spin direction can be never parallel to the field
without energy dissipation.

Then one has to take into consideration the effect of thermal
agitation, which can change the energy of the system and drive the
system into thermal equilibrium. Therefore, if $n_0$ departs from
its thermal equilibrium value $\bar{n}_0$, it will relaxes to
$\bar{n}_0$ exponentially with a characteristic time scale $T_R$,
called the relaxation time. Assuming that the relaxation velocity is
proportional to $\bar{n}_0-n_0$, $\partial\bar{n}_{0}/{\partial
t}\propto \bar{n}_0-n_0$, Eq. (\ref{eq:7a}) can be replaced with the
following one~\cite{Bloch},
\begin{eqnarray}
\frac{\partial}{\partial t}n_{0}= -2\sqrt{\alpha} n_0 (1-n_0)
     {\rm sin}\theta + \frac{\bar{n}_0-n_0}{T_R} ~.
\end{eqnarray}
$T_R$ scales qualitatively the thermal dissipation rate. Longer
$T_R$ denotes weaker thermal agitation.

Figure 3 displays the population of $M_F=0$ and $\pm1$ components,
as well as the magnetization $m$. The evolution of $n_0$ and $m$
reflects the dynamical process of the domain formation. As shown in
Fig. 3a and 3b, $n_0$ decreases {\it oscillatorily} with time driven
by the the thermal agitation, and meantime $m$ arises. A very
interesting result is that the oscillation amplitude {\it increases}
as the system relaxes. Generally, the thermal agitation suppresses
the macroscopic quantum coherence, and thus tends to kill
oscillations of the population, while here we see that the
oscillation is enhanced. Furthermore, the magnetization persists in
oscillating for a very long period of time after the amplitude
reaches its maximum. This result is qualitatively consistent with
the Berkeley group's observation of the unstable magnetization
mode\cite{Berkeley}.

If the thermal agitation gets stronger, the oscillation will be
enhanced first, and then suppressed, as shown in Fig. 3c and 3d.
Correspondingly, we divide the whole process into two periods with
respect to the domain formation, the growing period and the
stabilizing period. In the latter period, $n_0$ and $m$ oscillate
around their thermal equilibrium values and the amplitudes decrease
gradually, then we have a stable domain structure eventually. If
sketching the solution in the phase space diagram, one can find that
the growing period is represented by the trajectory in the libration
regime, and solutions for the stabilizing period lie in the rotation
regime. Based on this understanding, Fig. 3a and 3b show only the
growing period. Given the thermal agitation strong enough, the
quantum mechanical feature will be smeared out in both periods, as
Fig. 3e and 3f indicate.

According to the above discussions, the present model can
qualitatively describe the dynamic process of the domain formation
in a quenched $M_F=0$ condensate. Significantly, we show that the
spatial separation of magnetic domains brings about much nontrivial
effects on the spin dynamics of the ferromagnetic condensate. To get
a quantitative description, more local details of the particle
distribution and the relative phase should be considered.

In conclusion, we have investigated the dynamics of domain formation
in a ferromagnetic spinor Bose-Einstein condensate, taking into
account of the symmetry-breaking of the $M_F=1$ and $-1$ components.
Magnetic domains develop with the separation of $M_F=\pm1$
components. Our results suggest that the $M_F=0$ component in the
condensate can significantly decay to a very small value, far less
than $1/2$ as previous theories predicted. The domain structure is
formed and stabilized with the help of the thermal dissipation. A
thermally enhanced quantum-oscillation is observed during the
process.

This work is supported by the National Natural Science Foundation of
China (Grant No. 10504002), the Fok Yin-Tung Education Foundation,
China (Grant No. 101008), and the Ministry of Education of China
(NCET-05-0098). Q.G. acknowledges helpful discussions with K.
Sengstock, K. Bongs and L. You and support from the Deutsche
Forschungsgemeinschaft through the Graduiertenkolleg No. 463.

\end{document}